\documentclass[aps,preprint,showpacs,preprintnumbers,amsmath,amssymb]{revtex4}
\usepackage{amsmath,mathrsfs,amsbsy,color,graphicx,bm,amsthm,amsfonts}
\usepackage{units}
\usepackage{bbm}
\usepackage{times}
\usepackage{dcolumn}
\usepackage{mathrsfs}
\usepackage{amsmath,amssymb,epsfig}
\usepackage{amsmath}
\newcommand{\udots}{\mathinner{\mskip1mu\raise1pt\vbox{\kern7pt\hbox{.}}
\mskip2mu\raise4pt\hbox{.}\mskip2mu\raise7pt\hbox{.}\mskip1mu}}
\begin{document}

\title{ Quantum steering for different types of Bell-like states in gravitational background }
\author{Si-Han Li, Si-Han Shang, Shu-Min Wu\footnote{Email: smwu@lnnu.edu.cn (corresponding author)}}
\affiliation{ Department of Physics, Liaoning Normal University, Dalian 116029, China}


\begin{abstract}
In a relativistic framework, it is generally accepted that quantum steering of maximally entangled states provides greater advantages in practical applications compared to non-maximally entangled states.
In this paper, we investigate quantum steering  for four different types of Bell-like states of fermionic modes near the event horizon of a Schwarzschild black hole.
In some parameter spaces, the peak of steering asymmetry corresponds to a transition from two-way to one-way steerability for Bell-like states under the influence of the Hawking effect. It is intriguing to find that the fermionic steerability of the maximally entangled states experiences sudden death with the Hawking temperature, while the fermionic steerability of the non-maximally entangled states maintains indefinite persistence at infinite Hawking temperature. In contrast to prior research, this finding suggests that quantum steering of non-maximally entangled states is more advantageous than that  of maximally entangled states for processing quantum tasks in the gravitational background.  This surprising result overturns the traditional idea of ``the advantage of maximally entangled steering in the relativistic framework" and provides a new perspective for understanding the Hawking effect of the black hole.
\end{abstract}

\vspace*{0.5cm}
\maketitle
\section{Introduction}
In response to the seminal $1935$ paper published by Einstein, Podolsky and Rosen (EPR) \cite{Q1}, Schr\"{o}dinger \cite{Q2} introduced the concept of quantum steering.
EPR steering is a quantum phenomenon where one can manipulate the state of one subsystem by performing measurements on the other entangled subsystem \cite{Q3}.
Due to the fact that EPR steering requires quantum entanglement as a fundamental resource for steering remote states, while it is not always sufficient to violate Bell inequality, EPR steering can be perceived as a quantum correlation or a quantum resource that lies between quantum entanglement and Bell nonlocality.
There has been some advancement in investigating the intricate connections between Bell nonlocality, EPR steering and quantum entanglement \cite{Q4,Q5,Q6}, but this inquiry remains an unsolved mystery and an open question within the field.
Distinct from both quantum entanglement and Bell nonlocality, EPR steering possesses inherent asymmetric features. 
Compared with entanglement and Bell nonlocality, quantum steering is distinguished by its intrinsic asymmetry, which grants it unique advantages in quantum information tasks \cite{Q7,Q8,QQ8,Q9,Q10,Q11,Q12,Q13}. In particular, steering is the essential resource for one-sided device-independent quantum key distribution, where secure communication can be achieved even with untrusted devices \cite{Q7}. It also plays an important role in asymmetric quantum communication and quantum metrology, offering enhanced robustness against noise in practical implementations \cite{Q8,QQ8}.

The convergence of quantum information, quantum field theory, and gravity has given rise to the framework of relativistic quantum information. Understanding quantum phenomena in a relativistic context is crucial because gravitational effects on quantum systems must be considered in experimental scenarios over long distances.
The two primary research areas of relativistic quantum information involve utilizing quantum technology to explore the structure of spacetime and examining the influence of gravitational effects on quantum resources.
The field of quantum information in the gravitational background has obtained substantial interest in simulation, experiment, and theory.
In terms of simulation, the
Hawking radiation has been simulated by using a dc-SQUID array transmission line, a Bose-Einstein condensate, an optical analogue, and a superconducting circuit
\cite{Q14,Q15,Q16,Q17,Q18}.
From an experimental perspective
, Pan and his team have indicated that it was feasible to test the quantum effects in curved spacetime using the single-photon version of the Colella-Overhauser-Werner experiment \cite{Q19}.
Theoretically,
the influence of the gravitational effect on quantum steering, entanglement, discord, coherence, and entropic uncertainty relations has been investigated extensively \cite{ADG1,ADG2,ADG3,ADG4,ADG5,ADG6,ADG7,ADG8,ADG9,ADG10,ADG11,ADG12,ADG13,ADG14,ADG15,ADG16,ADG17,ADG18,ADG19,ADG20,Q20,Q21,Q22,Q23,ZQ23,Q24,Q25,Q26,Q28,Q29,Q30,QQ30,QQQ30,Q31,Q32,Q33,Q34,Q35,Q36,Q37,Q38,Q39,Q40,Q41,Q42,Q43,Q44,Q45,tQ45,WTL1,WTL2,WTL3,Q46}.
Previous studies have shown that the quantum resources of the maximally entangled states exhibit superior advantages compared to those of the non-maximally entangled states in the relativistic framework \cite{Q31,Q32,Q33,Q34,Q35,Q36,Q37,Q38,Q39,Q40,Q41,Q42,Q43,Q44,Q45,tQ45,WTL1,WTL2,WTL3,Q46,QV17,29QV17,fhk1,fhk2}.
However, preparing maximally entangled states is more challenging than preparing non-maximally entangled states in experiments. Therefore, non-maximally entangled states are commonly used to process quantum information in curved spacetime.
Based on the above facts, we naturally come up with a question: is it possible for  quantum steering of the non-maximally entangled states to be more advantageous than that of the maximally entangled states in curved spacetime? This question motivates our research. On the other hand, we aim to explore whether quantum steering in two directions has the same properties under the background of gravity, which serves as another motivation for our research.

Previous works \cite{Q40,Q41,Q42,Q43,Q44,Q45,tQ45,WTL1,WTL2,WTL3,Q46} generally hold that maximally entangled states possess optimal robustness for preserving entanglement in black hole backgrounds, and are thus regarded as the standard initial resource for relativistic quantum information processing. However, this study may present the first systematic demonstration that, under the influence of Hawking radiation, non-maximally entangled states exhibit greater robustness in quantum steering  across most parameter regimes. This result is significant in several respects: (1) it challenges the long-held assumption that maximally entangled states necessarily maintain an advantage in preserving quantum information in curved spacetime; (2) it reveals that the Hawking effect has a highly nontrivial influence on quantum correlations, particularly on quantum steering-a form of quantum correlation with intrinsic directionality;  (3) it offers a novel perspective for optimizing quantum resources in extreme gravitational environments; and (4) the black hole background can induce transitions among two-way steering, one-way steering, and even no-way steering, thereby revealing the asymmetry and nontrivial structure of quantum correlations in curved spacetime. These results not only deepen our understanding of the decoherence mechanisms induced by Hawking radiation, but also provide valuable theoretical guidance for the development of quantum communication protocols and quantum security schemes under strong gravitational fields \cite{Q9,Q10,Q11,Q12,Q13}.

In this paper, we investigate the impact of the Hawking effect on quantum steering for four distinct types of Bell-like states of fermionic fields shared by Alice and Bob, both positioned near the event horizon of a Schwarzschild black hole. We derive analytical expressions for fermionic steering related to these types of Bell-like states in curved spacetime. Our primary focus is on the influence of the Hawking effect on fermionic steering from Alice to Bob, from Bob to Alice, and the resulting steering asymmetry. Based on this model, three interesting conclusions can be obtained as: (i) quantum steering of non-maximally entangled states can be greater than quantum steering of maximally entangled states in the Schwarzschild black hole. In contrast, previous studies \cite{Q31,Q32,Q33,Q34,Q35,Q36,Q37,Q38,Q39,Q40,Q41,Q42,Q43,Q44,Q45,tQ45,WTL1,WTL2,WTL3,Q46,QV17,29QV17} have shown that the advantages of non-maximally entangled states are very obvious;
(ii) the change in  steering asymmetry displays the transition between two-way steering, one-way steering, and no-way steering in curved spacetime; (iii) selecting appropriate Bell-like states is crucial for effectively addressing relativistic quantum information tasks.

The structure of the paper is as follows.
In Sec.II, we briefly introduce the quantification of bipartite steering for the X-state.
In Sec.III, we discuss the quantization of Dirac field in the background of a Schwarzschild black hole.
In Sec.IV, we study fermionic steerability and its steering asymmetry for four different types of Bell-like states in Schwarzschild spacetime.
Finally, the conclusions are presented in Sec.V.

\section{Quantification of bipartite steering }
Quantum steering is a type of nonlocal correlation, which shows that one side of a bipartite quantum system can use local measurement to influence the quantum state of the other side.
In this paper, we consider the density matrix of the symmetric X-state, which can be expressed as
\begin{eqnarray}\label{w6}
\rho_x= \left(\!\!\begin{array}{cccc}
\rho_{11}&0&0&\rho_{14}\\
0&\rho_{22}&\rho_{23}&0\\
0&\rho_{32}&\rho_{33}&0\\
\rho_{41}&0&0&\rho_{44}
\end{array}\!\!\right),
\end{eqnarray}
where the real elements $\rho_{ij}$ fulfill $\rho_{ij}=\rho_{ji}$.
With regard to the X-state $\rho_x$ given by Eq.(\ref{w6}), the concurrence can be specifically denoted as \cite{LLL47}
\begin{eqnarray}\label{A7}
C(\rho_x)=2\max\{0, |\rho_{14}|-\sqrt{\rho_{22}\rho_{33}}, |\rho_{23}|-\sqrt{\rho_{11}\rho_{44}}\}.
\end{eqnarray}
For a two-qubit state $\rho_{AB}$ shared by Alice and Bob, the steering from Bob to Alice
can be witnessed, if the density matrix $\tau_{AB}^1$ defined as \cite{Q6,LLLL48}
\begin{eqnarray}\label{qq1}
\tau_{AB}^1=\frac{\rho_{AB}}{\sqrt{3}}+\frac{3-\sqrt{3}}{3}(\rho_A\otimes\frac{I}{2}),
\end{eqnarray}
is entangled, where $\rho_A$ is the reduced density matrix and $I$ indicates the two-dimension identity matrix.
Likewise, we can also witness the steering from Alice to Bob, if the state $\tau_{AB}^2$ defined as
\begin{eqnarray}\label{qq2}
\tau_{AB}^2=\frac{\rho_{AB}}{\sqrt{3}}+\frac{3-\sqrt{3}}{3}(\frac{I}{2}\otimes\rho_B),
\end{eqnarray}
is entangled, where $\rho_B=\rm{Tr}_A(\rho_{AB})$.
For the matrix of the X-state $\rho_x$, Eq.(\ref{qq1}) can be specifically written as
\begin{eqnarray}\label{qq3}
\tau_{AB}^{1,x}= \left(\!\!\begin{array}{cccc}
\frac{\sqrt{3}}{3}\rho_{11}+r&0&0&\frac{\sqrt{3}}{3}\rho_{14}\\
0&\frac{\sqrt{3}}{3}\rho_{22}+r&\frac{\sqrt{3}}{3}\rho_{23}&0\\
0&\frac{\sqrt{3}}{3}\rho_{23}&\frac{\sqrt{3}}{3}\rho_{33}+s&0\\
\frac{\sqrt{3}}{3}\rho_{14}&0&0&\frac{\sqrt{3}}{3}\rho_{44}+s
\end{array}\!\!\right),
\end{eqnarray}
with $r=\frac{(3-\sqrt{3})}{6}(\rho_{11}+\rho_{22})$ and $s=\frac{(3-\sqrt{3})}{6}(\rho_{33}+\rho_{44})$. From Eq.(\ref{A7}), it can be concluded that the state $\tau_{AB}^{1,x}$ is entangled, as long as one of the following inequalities
\begin{eqnarray}\label{qq5}
|\rho_{14}|^2>Q_a-Q_b,
\end{eqnarray}
or
\begin{eqnarray}\label{qqq5}
|\rho_{23}|^2>Q_c-Q_b,
\end{eqnarray}
is fulfilled, where
\begin{eqnarray}
 \nonumber&&Q_a=\frac{2-\sqrt{3}}{2}\rho_{11}\rho_{44}+\frac{2+\sqrt{3}}{2}\rho_{22}\rho_{33}
+\frac{1}{4}(\rho_{11}+\rho_{44})(\rho_{22}+\rho_{33}),\\ \nonumber
&&Q_b=\frac{1}{4}(\rho_{11}-\rho_{44})(\rho_{22}-\rho_{33}),\\ \nonumber
&& Q_c=\frac{2+\sqrt{3}}{2}\rho_{11}\rho_{44}+\frac{2-\sqrt{3}}{2}\rho_{22}\rho_{33}
+\frac{1}{4}(\rho_{11}+\rho_{44})(\rho_{22}+\rho_{33}) \nonumber.
\end{eqnarray}
The steering from Bob to Alice is thus witnessed.
Similarly, the steering from Alice to Bob can be witnessed by satisfying one of the inequalities,
\begin{eqnarray}\label{qq6}
|\rho_{14}|^2>Q_a+Q_b,
\end{eqnarray}
or
\begin{eqnarray}\label{qqq6}
|\rho_{23}|^2>Q_c+Q_b.
\end{eqnarray}

According to the inequality, we introduce $S^{A\rightarrow B}$ and $S^{B\rightarrow A}$ to quantify the steerability from Alice to Bob and from Bob to Alice, respectively.
Based on Eqs.(\ref{qq5})-(\ref{qqq6}), the steering from Alice to Bob $S^{A\rightarrow B}$ is found to be
\begin{eqnarray}\label{qq8}
S^{A\rightarrow B}={\rm{max}}\bigg\{0,\frac{8}{\sqrt{3}}(|\rho_{14}|^2-Q_a-Q_b),
\frac{8}{\sqrt{3}}(|\rho_{23}|^2-Q_c-Q_b)\bigg\}.
\end{eqnarray}
In the same way, the steering from Bob to Alice $S^{B\rightarrow A}$ can be written as
\begin{eqnarray}\label{qq7}
S^{B\rightarrow A}={\rm{max}}\bigg\{0,\frac{8}{\sqrt{3}}(|\rho_{14}|^2-Q_a+Q_b),
\frac{8}{\sqrt{3}}(|\rho_{23}|^2-Q_c+Q_b)\bigg\}.
\end{eqnarray}
Defining the coefficient as $\frac{8}{\sqrt{3}}$ assures that the steerability of the maximally entangled state is equal to $1$.

Different from quantum entanglement, quantum steering is essentially asymmetric, which means that $S^{A\rightarrow B}$ is not equal to $S^{B\rightarrow A}$.
Generally, the quantum steering can be categorized into three cases:
(i) no-way steering $S^{A\rightarrow B}= S^{B\rightarrow A}=0$, which cannot be manipulated in any direction;
(ii) two-way steering $S^{A\rightarrow B}>0$ and $ S^{B\rightarrow A}>0$, showing that the state can be manipulated in both directions;
(iii) one-way steering $S^{A\rightarrow B}>0$ and $ S^{B\rightarrow A}=0$, or vice versa $S^{B\rightarrow A}>0$ and $ S^{A\rightarrow B}=0$, meaning that the state can be manipulated in only one direction.
Therefore, quantum steering has richer characteristics compared with quantum entanglement.
Wherein, the last case is consistent with the asymmetric nature of quantum steering.
In order to measure the difference between the steering from $A$ to $B$ and from $B$ to $A$, the steering asymmetry is introduced, which can be defined as
\begin{eqnarray}\label{w10}
S_{AB}^\Delta&=&|S^{A\rightarrow B}-S^{B\rightarrow A}|.
\end{eqnarray}

\section{Quantization of Dirac field in Schwarzschild spacetime }

The metric of the Schwarzschild black hole \cite{Q46} is given by
\begin{eqnarray}\label{w11}
ds^2&=&-(1-\frac{2M}{r}) dt^2+(1-\frac{2M}{r})^{-1} dr^2\nonumber\\&&+r^2(d\theta^2
+\sin^2\theta d\varphi^2),
\end{eqnarray}
where the parameters $r$ and $M$ represent the radius and mass of the Schwarzschild black hole, respectively.
For simplicity, $\hbar, G, c$ and $k$ can be considered as unity. In Schwarzschild spacetime, the massless Dirac equation $[\gamma^a e_a{}^\mu(\partial_\mu+\Gamma_\mu)]\Phi=0$ \cite{Q49} can be expanded as
\begin{eqnarray}\label{w12}
&&-\frac{\gamma_0}{\sqrt{1-\frac{2M}{r}}}\frac{\partial \Phi}{\partial t}+\gamma_1\sqrt{1-\frac{2M}{r}}\bigg[\frac{\partial}{\partial r}+\frac{1}{r}+\frac{M}{2r(r-2M)} \bigg]\Phi \nonumber\\
&&+\frac{\gamma_2}{r}(\frac{\partial}{\partial \theta}+\frac{\cot \theta}{2})\Phi+\frac{\gamma_3}{r\sin\theta}\frac{\partial\Phi}{\partial\varphi}=0,
\end{eqnarray}
where $\gamma_i$ ($i=0,1,2,3$) are the Dirac matrices \cite{Q50, Q51}.

By solving the Dirac equation near the event horizon, a set of positive (fermions) frequency outgoing solutions for the regions inside and outside the event horizon can be derived as
\begin{eqnarray}\label{w13}
\Phi^+_{{\bold k},{\rm in}}\sim \phi(r) e^{i\omega u},
\end{eqnarray}
\begin{eqnarray}\label{w14}
\Phi^+_{{\bold k},{\rm out}}\sim \phi(r) e^{-i\omega u},
\end{eqnarray}
where $\phi(r)$ is the four-component Dirac spinor and $u=t-r_{*}$ with the tortoise coordinate $r_{*}=r+2M\ln\frac{r-2M}{2M}$ \cite{Q50,Q51}.
Here, $\bold k$ and $\omega$ indicate the wave vector and frequency, respectively, which fulfill $|\mathbf{k}|=\omega$ in the massless Dirac field.
According to Eqs.(\ref{w13}) and (\ref{w14}), the Dirac field $\Phi$ can be expanded as
\begin{eqnarray}\label{w15}
\Phi&=&\int
d\bold k[\hat{a}^{\rm in}_{\bold k}\Phi^{+}_{{\bold k},\text{in}}
+\hat{b}^{\rm in\dag}_{\bold k}
\Phi^{-}_{{\bold k},\text{in}}\nonumber\\ &+&\hat{a}^{\rm out}_{\bold k}\Phi^{+}_{{\bold k},\text{out}}
+\hat{b}^{\rm out\dag}_{\bold k}\Phi^{-}_{{\bold k},\text{out}}],
\end{eqnarray}
where $\hat{a}^{\rm in}_{\bold k}$ and $\hat{b}^{\rm in\dag}_{\bold k}$ are the fermion annihilation and antifermion creation operators in the internal region of the event horizon, and $\hat{a}^{\rm out}_{\bold k}$ and $\hat{b}^{\rm out\dag}_{\bold k}$ are the fermion annihilation and antifermion creation operators outside the event horizon, respectively. These annihilation and creation operators fulfill the canonical anticommutation relations $\{\hat{a}^{\rm out}_{\mathbf{k}},\hat{a}^{\rm out\dagger}_{\mathbf{k'}}\}=
\{\hat{b}^{\rm in}_{\mathbf{k}},\hat{b}^{\rm in\dagger}_{\mathbf{k'}}\}
=\delta_{\mathbf{k}\mathbf{k'}}.$ Therefore, the Schwarzschild vacuum can be defined as $\hat{a}^{\rm in}_{\bold k}|0\rangle_S=\hat{a}^{\rm out}_{\bold k}|0\rangle_S=0$.
Note that the modes $\Phi^\pm_{{\bold k},{\rm in}}$ and $\Phi^\pm_{{\bold k},{\rm out}}$ are commonly referred to as Schwarzschild modes.

According to the suggestions of Damour and Ruffini \cite{Q52}, we can use a complete basis of positive energy modes (Kruskal models) to perform an analytical continuation of Eqs.(\ref{w13}) and (\ref{w14}),
\begin{eqnarray}\label{w16}
\Psi^+_{{\bold k},{\rm out}}=e^{-2\pi M\omega} \Phi^-_{{-\bold k},{\rm in}}+e^{2\pi M\omega}\Phi^+_{{\bold k},{\rm out}},
\end{eqnarray}
\begin{eqnarray}\label{w17}
\Psi^+_{{\bold k},{\rm in}}=e^{-2\pi M\omega} \Phi^-_{{-\bold k},{\rm out}}+e^{2\pi M\omega}\Phi^+_{{\bold k},{\rm in}}.
\end{eqnarray}
Likewise, the Kruskal modes can also be applied to expand the Dirac fields in Kruskal spacetime
\begin{eqnarray}\label{w18}
\Phi&=&\int
d\bold k [2\cosh(4\pi M\omega)]^{-\frac{1}{2}}
[\hat{c}^{\rm in}_{\bold k}\Psi^{+}_{{\bold k},\text{in}}
+\hat{d}^{\rm in\dag}_{\bold k}
\Psi^{-}_{{\bold k},\text{in}}\nonumber\\ &+&\hat{c}^{\rm out}_{\bold k}\Psi^{+}_{{\bold k},\text{out}}
+\hat{d}^{\rm out\dag}_{\bold k}\Psi^{-}_{{\bold k},\text{out}}],
\end{eqnarray}
where $\hat{c}^{\sigma}_{\bold k}$ and $\hat{d}^{\sigma\dag}_{\bold k}$ with $\sigma=(\rm in, \rm out)$ are the fermion annihilation operators and antifermion creation operators acting on the Kruskal vacuum.

Based on Eqs.(\ref{w15}) and (\ref{w18}), the different decompositions of the same Dirac field in Schwarzschild and Kruskal modes, respectively, lead to the Bogoliubov transformation between the Schwarzschild and Kruskal operators,
\begin{eqnarray}\label{w19}
\hat{c}^{\rm out}_{\bold k}=\frac{1}{\sqrt{e^{-8\pi M\omega}+1}}\hat{a}^{\rm out}_{\bold k}-\frac{1}{\sqrt{e^{8\pi M\omega}+1}}\hat{b}^{\rm in\dag}_{\bold k},
\end{eqnarray}
\begin{eqnarray}\label{ww19}
\hat{c}^{\rm out\dag}_{\bold k}=\frac{1}{\sqrt{e^{-8\pi M\omega}+1}}\hat{a}^{\rm out\dag}_{\bold k}-\frac{1}{\sqrt{e^{8\pi M\omega}+1}}\hat{b}^{\rm in}_{\bold k}.
\end{eqnarray}
On the basis of the Bogliubov transformation given by Eqs.(\ref{w19}) and (\ref{ww19}), the Kruskal vacuum and excited states can be denoted in the Schwarzschild Fock space as
\begin{eqnarray}\label{w20}
\nonumber |0\rangle_K&=&\frac{1}{\sqrt{e^{-\frac{\omega}{T}}+1}}|0\rangle_{\rm out} |0\rangle_{\rm in}+\frac{1}{\sqrt{e^{\frac{\omega}{T}}+1}}|1\rangle_{\rm out} |1\rangle_{\rm in},\\
|1\rangle_K&=&|1\rangle_{\rm out} |0\rangle_{\rm in},
\end{eqnarray}
where $T=\frac{1}{8\pi M}$ indicates the Hawking temperature, $\{|n\rangle_{\rm out}\}$ and $\{|n\rangle_{\rm in}\}$ correspond to  the Schwarzschild number states for fermions outside the event horizon and antifermions inside respectively.

The Schwarzschild observer hovers in the external of the event horizon, and its Hawking radiation spectrum is given by \cite{Q51}
\begin{eqnarray}\label{w21}
N_F=\sideset{_K}{}{\mathop{\langle}}0|\hat{a}^{\rm out\dag}_{\bold k}\hat{a}^{\rm out}_{\bold k}|0\rangle_K=\frac{1}{e^{\frac{\omega}{T}}+1}.
\end{eqnarray}
Eq.(\ref{w21}) indicates that the Kruskal vacuum observed by the Schwarzschild observer, would be detected as a number of generated fermions $N_F$.
That is to say, the Schwarzschild observer in the exterior of the black hole can detect a thermal Fermi-Dirac statistic.

\section{Fermionic steering for different types of bell-like states in Schwarzschild spacetime }
The four different types of Bell-like states of the entangled fermionic modes in the asymptotically flat region of a Schwarzschild black hole can be described as
\begin{eqnarray}\label{w22}
|\phi^{{1},{\pm}}_{AB}\rangle=\gamma|0_{A}\rangle|0_{B}\rangle \pm \sqrt{1-\gamma^{2}}|1_{A}\rangle|1_{B}\rangle,
\end{eqnarray}
\begin{eqnarray}\label{w23}
|\Psi^{{2},{\pm}}_{AB}\rangle=\gamma|0_{A}\rangle|1_{B}\rangle \pm \sqrt{1-\gamma^{2}}|1_{A}\rangle|0_{B}\rangle,
\end{eqnarray}
where the subscripts $A$ and $B$ denote the modes associated with Alice and Bob, respectively.
Subsequently, both Alice and Bob are positioned outside the event horizon of the black hole.
Therefore, Alice and Bob will detect the thermal Fermi-Dirac statistics, and their detectors are found to be excited.
Employing Eq.(\ref{w20}), we can rewrite Eqs.(\ref{w22}) and (\ref{w23}) using Schwarzschild modes for both Alice and Bob as
\begin{eqnarray}\label{w24}
|\phi^{{1},{\pm}}_{A{\bar A}B{\bar B}}\rangle&=&
\gamma(
\cos \alpha\cos \beta|0\rangle_{A}|0\rangle_{\bar A}|0\rangle_{B}|0\rangle_{\bar B}
+\cos \alpha\sin \beta|0\rangle_{A}|0\rangle_{\bar A}|1\rangle_{B}|1\rangle_{\bar B}\nonumber\\&&
+\sin \alpha\cos \beta|1\rangle_{A}|1\rangle_{\bar A}|0\rangle_{B}|0\rangle_{\bar B}
+\sin \alpha\sin \beta|1\rangle_{A}|1\rangle_{\bar A}|1\rangle_{B}|1\rangle_{\bar B})\nonumber\\&&\pm
\sqrt{1-\gamma^{2}}|1\rangle_{A}|0\rangle_{\bar A}|1\rangle_{B}|0\rangle_{\bar B},
\end{eqnarray}
\begin{eqnarray}\label{w25}
|\Psi^{{2},{\pm}}_{A{\bar A}B{\bar B}}\rangle&=&
\gamma \cos \alpha|0\rangle_{A}|0\rangle_{\bar A}|1\rangle_{B}|0\rangle_{\bar B}
+\gamma \sin \alpha|1\rangle_{A}|1\rangle_{\bar A}|1\rangle_{B}|0\rangle_{\bar B}\nonumber\\&&\pm
\sqrt{1-\gamma^{2}}\cos \beta|1\rangle_{A}|0\rangle_{\bar A}|0\rangle_{B}|0\rangle_{\bar B}\pm
\sqrt{1-\gamma^{2}}\sin \beta|1\rangle_{A}|0\rangle_{\bar A}|1\rangle_{B}|1\rangle_{\bar B},
\end{eqnarray}
where the modes $\bar A$ and $\bar B$ are observed by hypothetical observers Anti-Alice and Anti-Bob inside the event horizon of the black hole, respectively.
Here, we define $\cos \alpha=\frac{1}{\sqrt{e^{-\frac{\omega_{A}}{T}}+1}}$,
$\sin \alpha=\frac{1}{\sqrt{e^{\frac{\omega_{A}}{T}}+1}}$,
$\cos \beta=\frac{1}{\sqrt{e^{-\frac{\omega_{B}}{T}}+1}}$,
and $\sin \beta=\frac{1}{\sqrt{e^{\frac{\omega_{B}}{T}}+1}}$ for simplicity,
where $\omega_{A}$ and $\omega_{B}$ are the frequencies of modes $A$ and $B$, respectively.

Since the exterior region of the black hole is causally disconnected from its interior, Alice and Bob cannot detect the physically inaccessible modes $\bar A$ and $\bar B$.
After tracing over these inaccessible modes, we can obtain the density matrix for Alice and Bob as
\begin{eqnarray}\label{w27}
 \rho^{{1},{\pm}}_{AB}=\left(\!\!\begin{array}{cccccccc}
\gamma^{2} \cos^{2}\alpha \cos^{2}\beta&0&0&\pm\gamma\sqrt{1-\gamma^{2}} \cos \alpha \cos \beta\\
0&\gamma^{2} \cos^{2}\alpha \sin^{2}\beta&0&0\\
0&0&\gamma^{2} \sin^{2}\alpha \cos^{2}\beta&0\\
\pm\gamma\sqrt{1-\gamma^{2}}\cos \alpha \cos \beta&0&0&\gamma^{2} \sin^{2}\alpha \sin^{2}\beta+1-\gamma^{2}
\end{array}\!\!\right),
\end{eqnarray}
\begin{eqnarray}\label{w28}
 \rho^{{2},{\pm}}_{AB}=\left(\!\!\begin{array}{cccccccc}
0&0&0&0\\
0&\gamma^{2} \cos^{2}\alpha&\pm\gamma\sqrt{1-\gamma^{2}}\cos \alpha \cos \beta&0\\
0&\pm\gamma\sqrt{1-\gamma^{2}}\cos \alpha \cos \beta&(1-\gamma^{2}) \cos^{2}\beta&0\\
0&0&0&\gamma^{2} \sin^{2}\alpha+(1-\gamma^{2})\sin^{2}\beta
\end{array}\!\!\right).
\end{eqnarray}

Based on the quantification of quantum steering, we can obtain the analytical expressions of quantum steering for $ \rho^{{1},{\pm}}_{AB}$ and $ \rho^{{2},{\pm}}_{AB}$ as
\begin{eqnarray}\label{w29}
S^{A\rightarrow B}_{{1},{\pm}}&=&\max\bigg\{0,
\frac{4}{\sqrt{3}}\cos^{2}\beta [
-\gamma^{4}\sin^{2}\beta -\gamma^{2}(1-\gamma^{2})\sin^{2}\alpha
+\sqrt{3}\gamma^{2}(1-\gamma^{2})\cos^{2}\alpha \nonumber\\
&&-2\gamma^{4}\cos^{2}\alpha\sin^{2}\alpha\sin^{2}\beta]\bigg\},
\end{eqnarray}
\begin{eqnarray}\label{w30}
S^{B\rightarrow A}_{{1},{\pm}}&=&\max\bigg\{0,
\frac{4}{\sqrt{3}}\cos^{2}\alpha[
-\gamma^{4}\sin^{2}\alpha-\gamma^{2}(1-\gamma^{2})\sin^{2}\beta
+\sqrt{3}\gamma^{2}(1-\gamma^{2})\cos^{2}\beta \nonumber\\
&&-2\gamma^{4}\cos^{2}\beta\sin^{2}\alpha\sin^{2}\beta
]\bigg\},
\end{eqnarray}
\begin{eqnarray}\label{w31}
S^{A\rightarrow B}_{{2},{\pm}}&=&\max\left\{0,
\frac{4}{\sqrt{3}}(1-\gamma^{2})\cos^{2}\beta[
-\gamma^{2}\sin^{2}\alpha
-(1-\gamma^{2})\sin^{2}\beta
+\sqrt{3}\gamma^{2}\cos^{2}\alpha
]\right\},
\end{eqnarray}
\begin{eqnarray}\label{w32}
S^{B\rightarrow A}_{{2},{\pm}}&=&\max\left\{0,
\frac{4}{\sqrt{3}}\gamma^{2}\cos^{2}\alpha[
-\gamma^{2}\sin^{2}\alpha
-(1-\gamma^{2})\sin^{2}\beta
+\sqrt{3}(1-\gamma^{2})\cos^{2}\beta
]\right\}.
\end{eqnarray}
From the above formula, we observe that the steerability is related to the Hawking temperature $T$.
In other words, the Hawking radiation of the black hole affects the fermionic steerability. In Minkowski spacetime, the steerability from Alice to Bob is equal to the steerability from Bob to Alice. However, in curved spacetime, the influence of Hawking radiation causes the steerability from Alice to Bob to differ from the steerability from  Bob to Alice. To assess the degree of steerability asymmetry in curved spacetime, we obtain their expressions as
\begin{eqnarray}\label{w33}
S_{{1},{\pm}}^\Delta&=&\mid S^{A\rightarrow B}_{{1},{\pm}}-S^{B\rightarrow A}_{{1},{\pm}}\mid,
\end{eqnarray}
\begin{eqnarray}\label{w34}
S_{{2},{\pm}}^\Delta&=&\mid S^{A\rightarrow B}_{{2},{\pm}}-S^{B\rightarrow A}_{{2},{\pm}}\mid.
\end{eqnarray}

\begin{figure}
\begin{minipage}[t]{0.5\linewidth}
\centering
\includegraphics[width=3.0in,height=5.2cm]{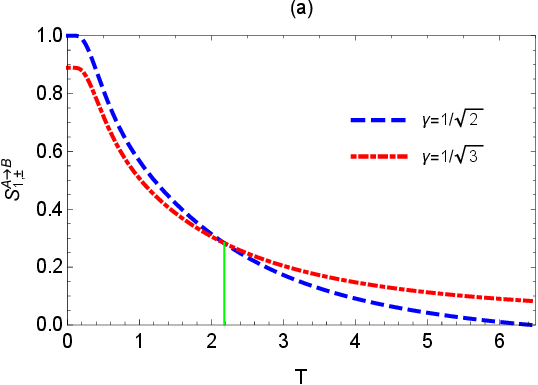}
\label{fig1a}
\end{minipage}%
\begin{minipage}[t]{0.5\linewidth}
\centering
\includegraphics[width=3.0in,height=5.2cm]{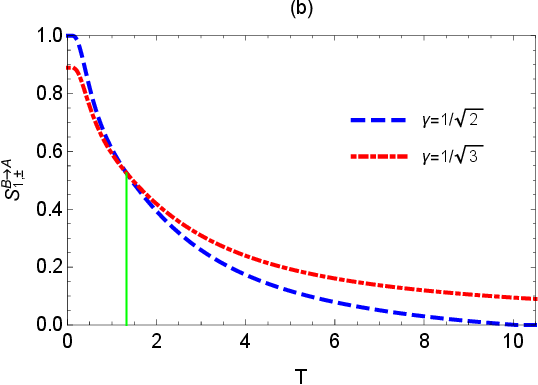}
\label{fig1b}
\end{minipage}%

\begin{minipage}[t]{0.5\linewidth}
\centering
\includegraphics[width=3.0in,height=5.2cm]{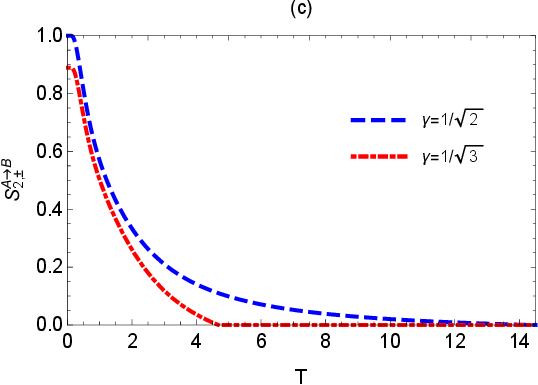}
\label{fig1c}
\end{minipage}%
\begin{minipage}[t]{0.5\linewidth}
\centering
\includegraphics[width=3.0in,height=5.2cm]{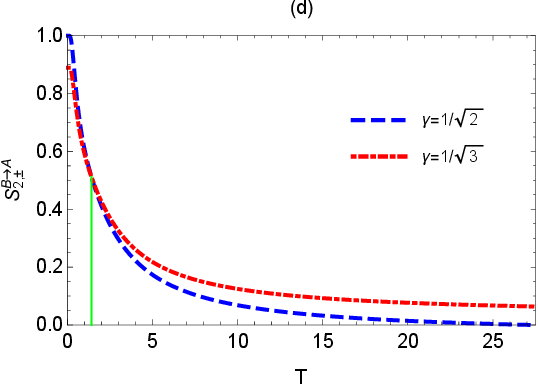}
\label{fig1d}
\end{minipage}%
\caption{The fermionic steerability between Alice and Bob as functions of the Hawking temperature $T$ for  different initial parameters  $\gamma$ with fixed $\omega_{A}=1$ and $\omega_{B}=5$. }
\label{Fig1}
\end{figure}

\begin{figure}
\begin{minipage}[t]{0.5\linewidth}
\centering
\includegraphics[width=3.0in,height=5.2cm]{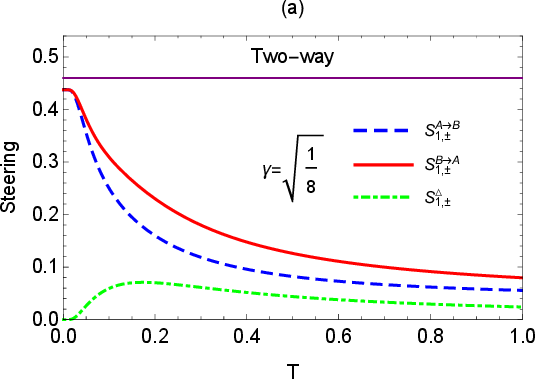}
\label{fig2a}
\end{minipage}%
\begin{minipage}[t]{0.5\linewidth}
\centering
\includegraphics[width=3.0in,height=5.2cm]{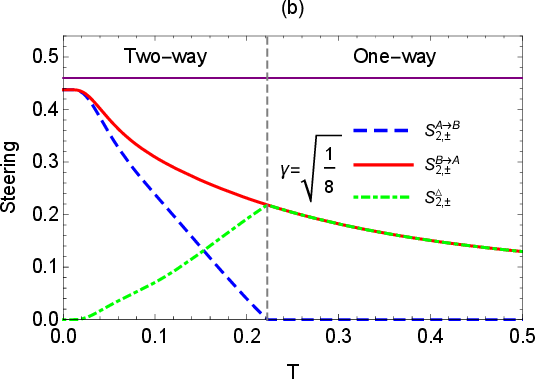}
\label{fig2b}
\end{minipage}%

\begin{minipage}[t]{0.5\linewidth}
\centering
\includegraphics[width=3.0in,height=5.2cm]{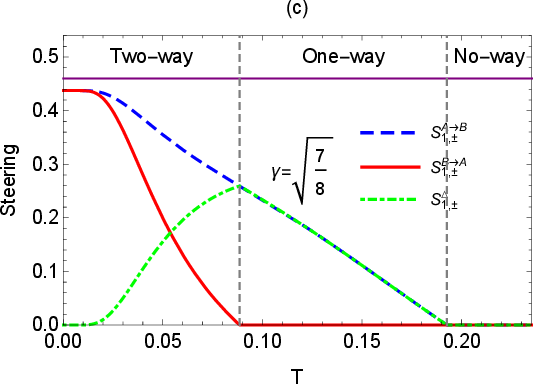}
\label{fig2c}
\end{minipage}%
\begin{minipage}[t]{0.5\linewidth}
\centering
\includegraphics[width=3.0in,height=5.2cm]{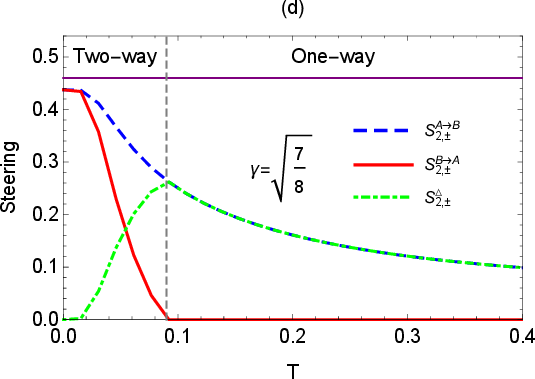}
\label{fig2d}
\end{minipage}%

\begin{minipage}[t]{0.5\linewidth}
\centering
\includegraphics[width=3.0in,height=5.2cm]{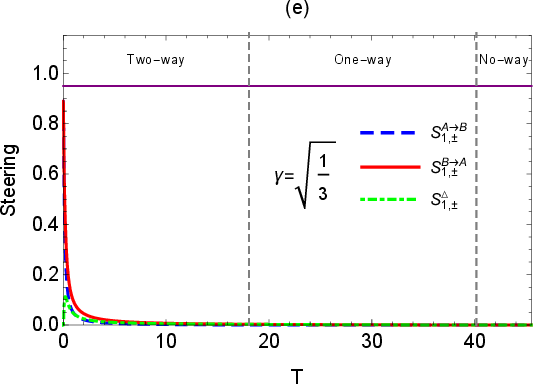}
\label{fig2e}
\end{minipage}%
\begin{minipage}[t]{0.5\linewidth}
\centering
\includegraphics[width=3.0in,height=5.2cm]{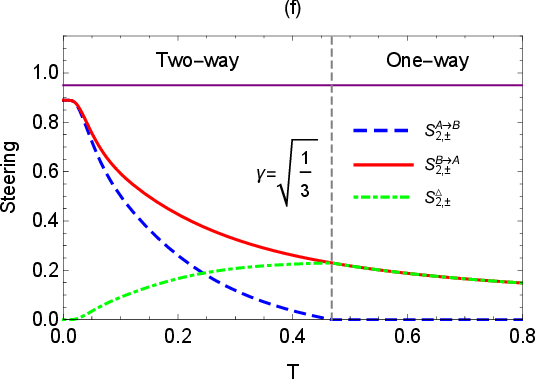}
\label{fig2f}
\end{minipage}%

\begin{minipage}[t]{0.5\linewidth}
\centering
\includegraphics[width=3.0in,height=5.2cm]{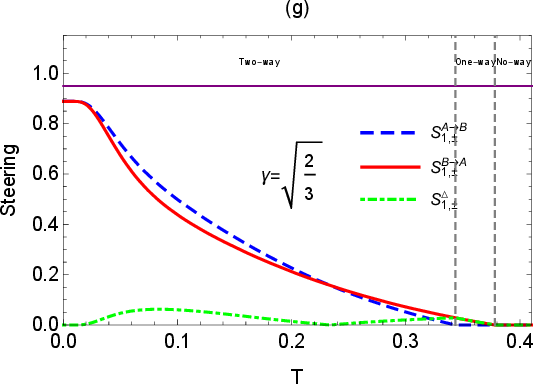}
\label{fig2g}
\end{minipage}%
\begin{minipage}[t]{0.5\linewidth}
\centering
\includegraphics[width=3.0in,height=5.2cm]{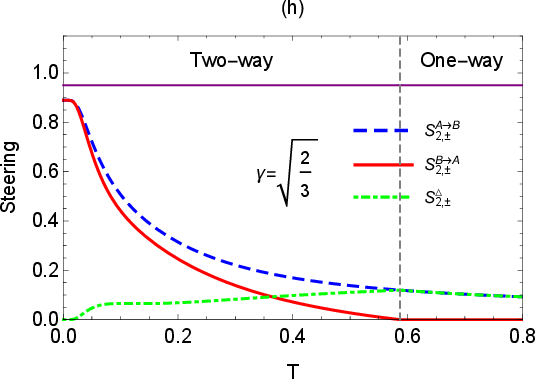}
\label{fig2h}
\end{minipage}%
\caption{The fermionic steerability and steering asymmetry between Alice and Bob as functions of the Hawking temperature $T$ for  different initial parameters  $\gamma$ with  fixed $\omega_{A}=0.1$ and $\omega_{B}=0.5$. }
\label{Fig2}
\end{figure}

In Fig.\ref{Fig1}, we plot the fermionic steerability from $A$ to $B$ and  from $B$ to $A$ with the Hawking temperature $T$ for different initial parameters  $\gamma$ and four different types of Bell-like states.
It is shown that  as the Hawking temperature $T$ increases, the fermionic steerability first decreases and then suffers sudden death or approaches an asymptotic value, depending on the selection of the initial state. From Fig.\ref{Fig1}(a) and (b), we can see that for the Bell-like states $|\phi^{{1},{\pm}}_{AB}\rangle$, with the increase of Hawking temperature $T$, the fermionic steerability of the maximally entangled states ($\gamma=\frac{1}{\sqrt{2}}$) initially exceeds that of the non-maximally entangled states ($\gamma=\frac{1}{\sqrt{3}}$). However, as the temperature continues to rise, the fermionic steerability of the non-maximally entangled states eventually surpasses that of the maximally entangled states in Schwarzschild spacetime.
From Fig.\ref{Fig1}(c) and (d), we also see that for the Bell-like states $|\phi^{{2},{\pm}}_{AB}\rangle$, the fermionic steerability $S^{A\rightarrow B}$ of the maximally entangled states remains consistently greater than that of the non-maximally entangled states ($\gamma=\frac{1}{\sqrt{3}}$), while the fermionic steerability $S^{B\rightarrow A}$ of the non-maximally entangled states can surpass that of the maximally entangled states in curved spacetime. This indicates that the fermionic steerability of the maximally entangled states undergoes sudden death under the influence of the Hawking effect, while the fermionic steerability of the non-maximally entangled states can persist indefinitely. This finding contrasts with the previous conclusion that maximally entangled states consistently hold an advantage over non-maximally entangled states in a relativistic setting \cite{Q31,Q32,Q33,Q34,Q35,Q36,Q37,Q38,Q39,Q40,Q41,Q42,Q43,Q44,Q45,tQ45,WTL1,WTL2,WTL3,Q46,QV17,29QV17}. As a result, it may be more advantageous to select non-maximally entangled states for handling relativistic quantum information tasks in a strong gravitational environment.

In Fig.\ref{Fig2}, we plot the fermionic steerability from $A$ to $B$, the fermionic steerability from $B$ to $A$, and steering asymmetry with the Hawking temperature $T$ for  different initial parameters  $\gamma$.  We observe that even if the initial states have the same amount of initial steering, the degradation changes under swapping $\gamma$ and $\sqrt{1-\gamma^{2}}$ in curved spacetime. Recall that our initial entangled states, given in Eqs.(\ref{w22}) and (\ref{w23}), have the initial steerability $S^{A\rightarrow B}=S^{B\rightarrow A}=4\gamma^2(1-\gamma^2)$ that does not change under $\gamma\rightarrow\sqrt{1-\gamma^{2}}$. It is shown that quantum steering is symmetric in the asymptotically flat region, while the Hawking effect of the black hole destroys the steering symmetry. In certain scenarios, we observe that the initial system undergoes a transformation from two-way steering to one-way steering, and subsequently transitions from one-way steering to no-way steering with the growth of the Hawking temperature $T$.
Interestingly, in certain parameter spaces, the peak of steering asymmetry signifies that the system is undergoing a transition from two-way steering to one-way steering. In other words, the attainment of maximal steering asymmetry marks the point of transition between one-way and two-way steerability for Bell-like states  under the influence of the Hawking effect.

In our analysis, the asymmetry of quantum steering in Schwarzschild spacetime arises from the inequivalent influence of Hawking radiation on the fermionic modes associated with Alice and Bob. While steering is symmetric in flat spacetime, the presence of the event horizon introduces thermal noise and information loss, which differently affect the two subsystems and thus break the symmetry. Physically, this asymmetry reveals that quantum correlations in curved spacetime are directionally vulnerable: the ability of one party to steer the other may persist, vanish, or undergo transitions between two-way, one-way, and no-way steering, depending on the gravitational conditions. Such directional behavior not only provides insight into the decoherence mechanisms induced by black holes but also has practical implications for selecting the optimal direction of steering in relativistic quantum communication protocols.

\section{Conclusions}
In this paper, we have explored how the Hawking effect influences fermionic  steering, concentrating on four distinct types of Bell-like states in Schwarzschild spacetime. Our model involves two subsystems, $A$ and $B$, located outside the black hole's event horizon and observed by Alice and Bob, respectively.
We find that the fermionic steerability of maximally entangled states undergoes sudden death as a result of the Hawking effect of the black hole. In contrast, the fermionic steerability of non-maximally entangled states can persist indefinitely with the Hawking temperature. This suggests that using quantum steering of non-maximally entangled states offers greater advantages for relativistic quantum information tasks. This result contrasts with previous research, which consistently showed that quantum resources in maximally entangled states outperform those in non-maximally entangled states in relativistic contexts \cite{Q31,Q32,Q33,Q34,Q35,Q36,Q37,Q38,Q39,Q40,Q41,Q42,Q43,Q44,Q45,tQ45,WTL1,WTL2,WTL3,Q46,QV17,29QV17}.
It is shown that, although the initial states possess the same amount of initial steering, the degradation behaves differently when $\gamma$ and $\sqrt{1-\gamma^{2}}$ are swapped in curved spacetime. Additionally, quantum steering exhibits symmetry in the asymptotically flat region, but this symmetry can be disrupted by the Hawking effect of the black hole. In certain parameter spaces,  the maximum of steering asymmetry indicates that Bell-like states transition from two-way  to one-way steering under the influence of the Hawking effect. These conclusions overturn the conventional idea of ``the advantage of quantum steering of maximally entangled states" and  can guide us in selecting appropriate quantum states for quantum steering in handling relativistic quantum information tasks.

\textbf{ Data availability}

No data was used for the research described in the article.

\textbf{Declaration of competing interest}

The authors declare that they have no known competing financial
interests or personal relationships that could have appeared to influence
the work reported in this paper.

\begin{acknowledgments}
This work is supported by the National Natural
Science Foundation of China (Grant Nos. 12205133), LJKQZ20222315, JYTMS20231051, and  the Special Fund for Basic Scientific Research of Provincial Universities in Liaoning under grant NO. LS2024Q002.	
\end{acknowledgments}


\end{document}